\documentclass[twocolumn,preprintnumbers,amsmath,amssymb,showpacs]{revtex4}
\usepackage{graphicx}
\usepackage{dcolumn}
\usepackage{bm}
\usepackage{ulem}

\begin{document}

\title{Non-local double-path Casimir phase in atom interferometers}

\author{Fran\c{c}ois Impens$^{1}$, Ryan O. Behunin$^{2,3}$, Claudio Ccapa Ttira$^{4}$, and Paulo A. Maia Neto$^{4}$}

\affiliation{$^{1}$ Observatoire de la C\^{o}te d'Azur (ARTEMIS), Universit\'e de Nice-Sophia Antipolis, CNRS, 06304 Nice, France} 
\affiliation{$^{2}$ Theoretical Division, MS B213, Los Alamos National Laboratory, Los Alamos, NM 87545, USA}
\affiliation{$^{3}$ Center for Nonlinear Studies}
\affiliation{$^{4}$ Instituto de F\'{i}sica, Universidade Federal do Rio de Janeiro,  Rio de Janeiro, RJ 21941-972, Brazil}

\date{\today }

\pacs{03.65.Yz} 
\pacs{42.50.Ct} 
\pacs{03.75.Dg} 

\begin{abstract}

We present an open quantum system theory of atom interferometers evolving in the quantized electromagnetic field bounded by an ideal conductor. Our treatment reveals an unprecedented feature of matter-wave propagation, namely the appearance of a non-local double-path phase coherence. 
In the standard interpretation of interferometers, one associates well-defined separate phases to individual paths. Our non-local phase coherence is instead associated to pairs of paths. It arises from the coarse-graining over the quantized electromagnetic field and internal atomic degrees of freedom, which play the role of a common reservoir for the pair of paths and lead to a non-Hamiltonian evolution of the atomic waves. We develop a diagrammatic interpretation  and estimate the non-local phase for realistic experimental parameters.
\end{abstract}

\maketitle

Atom interferometry~\cite{Cronin09} has become a field of great importance for both basic and applied science, enabling, in particular, the realization of extremely accurate inertial sensors~\cite{InertialSensors,Kasevich07}. With the advent of the coherent atomic waves guided on chips~\cite{Reichel01}, the investigation of atom-surface interactions has become a frontier for such systems. Already, atom interferometers have been used 
to probe the van der Waals regime~\cite{CroninVigue}. This experimental effort calls for a complete theory of atom interferometers in the 
presence of quantum fluctuations of the electromagnetic (EM) field.

In this letter, we layout such theory for a beam of neutral atoms and find an unusual new concept in interferometry: a non-local phase associated to pairs of paths rather than to individual ones. First, we present a theory of atomic phase-shifts 
 taking the effect of field and atomic dipole fluctuations separately over each interferometer arm. This method already contains novel dynamical corrections, which cannot be obtained by standard techniques suitable for atoms driven by conservative forces. However, it neglects quantum correlations, mediated by the  field, between the atomic wave-packets evolving along the separate arms.  In order to capture this effect, we develop a theory of atom interferometers based on the influence functional method~\cite{FeynmanVernon}, which allows us to derive the non-Hamiltonian evolution of the external atomic observables after coarse-graining over the  quantized electromagnetic field and internal atomic (dipole) degrees of freedom. 
 A non-local  phase shift arises as a consequence of 
the finite correlation time of dipole fluctuations interacting across a pair of interferometer paths. 
  It is absent in the standard Hamiltonian
   treatment of matter-wave dynamics with conservative forces, which shows that the effect of quantum vacuum and zero-point dipole fluctuations on atomic waves cannot be  understood as an effective potential.
 
 The influence functional method also allows one to consider the decoherence effect~\cite{CasimirDecoherence},
 another important consequence of the 
  non-unitary nature of the matter-wave dynamics. However, in this letter
   we focus on the non-local real phase shifts beyond the expected loss of contrast in the fringe pattern. Phase shifts induced by the environment were also
   considered in the context of geometrical phases for spin one-half systems~\cite{GeometricPhase}.

\begin{figure}[htbp]
\begin{center}
\includegraphics[width=8.5cm]{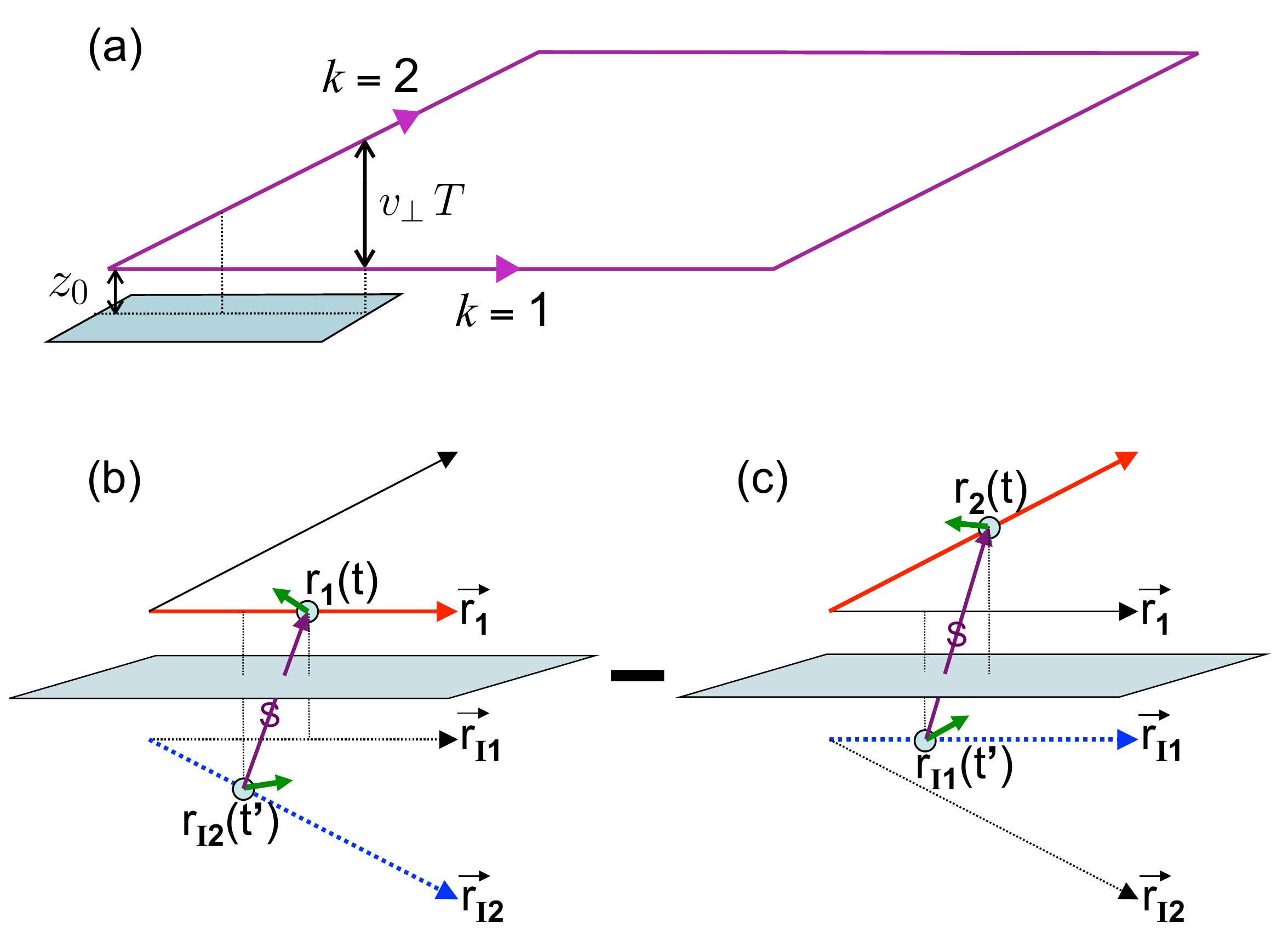}
\end{center}  \caption{(color
  online).
  (a) Atom interferometer flying nearby a conducting plate at $z=0$ during the time $T,$ with the arm $k=1$ parallel to the plate (distance $z_0$) 
  and the arm $k=2$ flying away with a normal velocity $v_{\perp}.$
 (b)-(c) Diagrammatic representation of the scattering double-path phase.
   Each diagram involves an advanced time $t$ on one path (red) and a retarded time $t'$ on the image of the other path (blue). The purple lines stand for the scattering retarded electric field Green's functions. The delay $\tau=t-t'$ is given by $|\mathbf{r}_k(t)-\mathbf{r}_l(t')|/c$  with $k,l=1,2, \: k \neq l$.
 The plate-dependent double-path phase shift $\phi^{\rm DP}_S$ results from the subtraction between diagrams (b) and (c) [see Eq.~(\ref{phiS})]. The cross-talk between the wave-packets is stronger in diagram (b), since it corresponds to a shorter interaction distance, thus resulting in a positive phase shift. 
   }
\label{fig:diagrams}
\end{figure}

 We consider 
 the Mach-Zehnder atom interferometer depicted in Fig.~\ref{fig:diagrams}(a), with
 two arms sharing the same origin, and followed by atoms flying above a metallic plate between the instants $t=0$ and $t=T$. 
 One of the arms is parallel to the plate, and the other has a velocity component $v_{\perp}$ perpendicular to the plate. We assume that the atomic motion is
 non-relativistic. 
For the typical atomic velocities used in van der Waals experiments \cite{CroninVigue}, the deflection of the average trajectory due to the van der Waals force is negligible. 
 The considered initial time is immediately after the first atomic beam-splitter: the initial external atomic quantum state is taken as a coherent superposition of two dilute Gaussian wave-packets of common average position $\mathbf{r}_{0}$ and different average momenta $\mathbf{p}_{0 k}$ for $k=1,2$. Here we focus on the effect of the plate, and thus calculate the phase accumulated between $t=0$ and $t=T.$

 Within the standard formalism of atom interferometers,
 the external atomic observables, namely the center-of-mass position 
 $\hat{\mathbf{r}}_a$ and momentum $\hat{\mathbf{p}}$, are supposed to evolve according to the Hamiltonian 
 $\hat{H}_E=\frac {\hat{\mathbf{p}}^2} {2 m}+ V(\hat{\mathbf{r}}_a)$. $V(\mathbf{r})$ is a quadratic and separable potential, 
 and $m$ is the atomic mass. 
 The internal atomic degrees of freedom (d.o.f.), initially in the ground state, follow a general Hamiltonian $\hat{H}_D$. The $ABCD$ theorem for atomic waves~\cite{BordeABCD} then
  shows that dilute Gaussian atomic wave-packets acquire a phase proportional to the action $S(\mathbf{r}_{0},\mathbf{p}_{0},t)$ along the classical trajectory:
$S(\mathbf{r}_{0},\mathbf{p}_{0},t)= \int_{0}^{t} \! dt' \left[  \frac {\mathbf{p}^2(t')}  {2 m} - \langle \hat{H}_D \rangle(t') - V\left(\mathbf{r}(t')\right) \right]$. Extensions for interacting samples~\cite{ABCDinteraction} in the paraxial approximation~\cite{ABCDatomlaser} have been recently developed.

Here we go beyond the standard $ABCD$ approach by taking into account quantum  dipole and EM fluctuations. 
 We first describe the effect of the quantum EM field  within  a local description of the interferometer. 
  We consider the full Hamiltonian $\hat{H}=\hat{H}_E+\hat{H}_D+\hat{H}_F+\hat{H}_{AF}$ with the field Hamiltonian $\hat{H}_F= \sum_{\lambda} \hbar \omega_{\lambda}  \left(\hat{a}_{\lambda}^{\dagger} \hat{a}_{\lambda} +1/2 \right)$ given as a sum over normal modes $\lambda$
(frequency $\omega_{\lambda}$,  annihilation and creation operators $\hat{a}_{\lambda}$ and $\hat{a}_{\lambda}^{\dagger}$).
    The interaction Hamiltonian in the electric dipole approximation is 
     $\hat{H}_{AF} = - \hat{\mathbf{d}} \cdot \hat{\mathbf{E}}(\hat{\mathbf{r}}_a)$
      ($\hat{\mathbf{d}}=$
     dipole operator). 
  We  assign to each interferometer path $k$ a phase corresponding to the average dipole interaction energy for the Gaussian atomic packet following this prescribed path, 
i.e. $\phi^{(k)}_{loc}= \frac {1} {4 \hbar} \int_{0}^{T} dt  \langle   \hat{\mathbf{d}} \cdot \hat{\mathbf{E}}(\hat{\mathbf{r}}_a)  + \hat{\mathbf{E}}(\hat{\mathbf{r}}_a)  \cdot \hat{\mathbf{d}} \rangle_k (t)$.  To derive this phase, we 
use  linear response theory and treat the dipole interaction Hamiltonian $H_{AF}$ as a perturbation~\cite{WylieSipe}. 
The phase is thus calculated to first order in the atomic polarizability. Our results hold as long as the atom-surface distance is much larger than  
   the typical atomic size, a condition  also necessary for the validity of the electric dipole approximation. 

 The phase $\phi^{(k)}_{loc}$  is then expressed as a propagation integral involving two contributions, namely the response of the atomic dipole to the on-atom electric field fluctuations and the response of the on-atom electric field to the dipole fluctuations. The fluctuations of the dipole and electric field are captured by Hadamard Green's functions, whereas the susceptibilities (polarizability for the atom) correspond to retarded Green's functions~\cite{AtomPhotons}. The retarded ($R$) and Hadamard ($H$)
  Green's functions read, respectively,
\begin{eqnarray}
\label{eq:def Green functions}
G^{R}_{\hat{\mathbf{O}}, \: ij}(x,x') & = & \frac {i} {\hbar} \theta(t-t') \langle [ \hat{O}_i^{f}(x),  \hat{O}_j^{f}(x')  ] \rangle\\
 G^{H}_{\hat{\mathbf{O}}, \: ij}(x,x') & = &  \frac 1 \hbar \langle  \{ \hat{O}_i^{f}(x),  \hat{O}_j^{f}(x')  \} \rangle \label{Hadamard}
\end{eqnarray}
with the operator $\hat{\mathbf{O}}=\hat{\mathbf{d}},\hat{\mathbf{E}}$, where $\hat{\mathbf{O}}^f$ corresponds to the free-evolving Heisenberg operator, and with the Cartesian indices $i,j=1,2,3$.
$\theta(t-t') $ denotes the Heaviside step function. 
 For the dipole operator $\hat{\mathbf{O}}=\hat{\mathbf{d}}$, the arguments of the Green's functions are two instants $(x,x') \equiv (t,t')$.  For the electric field operator $\hat{\mathbf{O}}=\hat{\mathbf{E}}$, these arguments are two four-vectors $(x,x') \equiv (\mathbf{r},t;\mathbf{r}',t')$. By isotropy of the free-evolving dipole operators,
 we have $G^{H,R}_{\hat{d},\: ij}(t,t')=G^{H,R}_{\hat{d}}(t,t') \,\delta_{ij}.$ Consequently, only the trace of the electric field Green's functions with respect to their tensorial components, noted from now on $\mathcal{G}_{\hat{\mathbf{E}}}^{R(H)}\left(x,x'\right)= \sum_{i}  G^{R(H)}_{\hat{\mathbf{E}} \: ii} \left(x,x'\right)$, shall contribute to the effects discussed hereafter.
  
 We assume that the width of the atomic wave-packet is small compared to the relevant EM field wavelengths, which allows us to replace the on-atom electric field operator at a given time by the electric field evaluated at the corresponding average atomic position $\mathbf{r}_k(t)$ on the considered path $k.$
One can then write the local Casimir phase as:
\begin{eqnarray}
& \: & \phi_{loc}^{(k)}  \simeq  \frac {1} {4}\int_{0}^T dt \int_{0}^T  dt' \left[ 
 \frac {} {}   G^H_{\hat{d}}(t,t') \mathcal{G}^R_{\hat{\mathbf{E}}} \left(r_k(t),r_k(t')\right) \right. \nonumber \\
&  & \qquad \left. \frac {} {} \! + G^R_{\hat{d}}(t,t')  \mathcal{G}^H_{\hat{\mathbf{E}}}\left(r_k(t),r_k(t')\right)
 \right] 
\label{eq:phase usual atom interferometry2}
\end{eqnarray}
with the four-vectors $r_k(t)\equiv(\mathbf{r}_{k}(t),t).$ The first term in (\ref{eq:phase usual atom interferometry2}) represents the contribution of dipole fluctuations modifying the on-atom 
electric field (radiation reaction). 
The second term accounts for the polarization of the atom by EM field fluctuations since
the retarded dipole Green's function 
$G^R_{\hat{d}}(t,t')$ represents the atomic linear polarizability in the time domain. 

We derive the electric field Green's functions defined by Eqs.~(\ref{eq:def Green functions})-(\ref{Hadamard})
 by taking the full normal mode decomposition of the electric field operator in the presence of a planar perfect conductor placed at $z=0.$ The Hadamard Green's function can be obtained from the retarded one thanks to the fluctuation-dissipation theorem, thus we focus on the latter. This function can be written as the sum of free-space and scattering contributions $\mathcal{G}^R_{\hat{\mathbf{E}}}(x,x')  = \mathcal{G}_{\hat{\mathbf{E}}}^{R, 0}(x,x') +\mathcal{G}_{\hat{\mathbf{E}}}^{R, S}(x,x')$ depending on the time difference $\tau=t-t'$ and on the positions $(\mathbf{r},\mathbf{r}')$. 
 Our result for the free-space contribution $\mathcal{G}_{\hat{\mathbf{E}}}^{R, 0}(x,x')$
  depends only on the relative distance $|\mathbf{r}-\mathbf{r}'|$ and is consistent with known expressions for the electric field commutators~\cite{Heitler}. 
 The scattering contribution $\mathcal{G}_{\hat{\mathbf{E}}}^{R, S}(x,x') $ depends on the propagation distance $|\mathbf{R}_{\rm I}|=
 |\mathbf{r}-\mathbf{r}'_{\rm I}|$ between the point $\mathbf{r}$ and the 
   image  $\mathbf{r}'_{\rm I}$  of the source point ${\bf r}'$ with respect to the plate:
\begin{eqnarray}
\label{G_sca}
 \mathcal{G}_{\hat{\mathbf{E}}}^{R, S}(x,x')   & = & \frac {\theta(\tau)} {2 \pi\epsilon_0} \frac {\partial^2} {\partial z \partial z'} \left( \frac {\delta ( \tau-|\mathbf{R}_{\rm {I}}|/c )} {|\mathbf{R}_{{\rm I}}|}  \right). 
\end{eqnarray}

 As an independent check,  we evaluate the local phase given by (\ref{eq:phase usual atom interferometry2}) for the two paths shown in Fig.~1(a).
 For the path $k=1$ parallel to the plate,  we find, in agreement with the $ABCD$ approach,
 the van der Waals phase $\phi_{loc}^{(1)}= -V_{\rm vdW}(z_0)\,T/\hbar $  for long interaction times $T,$ 
 where $V_{\rm vdW}(z_0)$ is the  van der Waals   (Casimir-Polder for long distances) potential at the atom-surface distance $z_0.$
 On the other hand, 
Eq.~(\ref{eq:phase usual atom interferometry2})
  already contains non-trivial dynamical Casimir effects 
\cite{Ryan2}, beyond the plain $ABCD$ integration of the van der Waals potential taken at the instantaneous position along the classical trajectory, particularly in the case of a time-dependent  atom-surface distance as in the trajectory ${\bf r}_2(t)$ shown in Fig.~1(a), for which (\ref{eq:phase usual atom interferometry2}) 
and (\ref{G_sca}) lead to a dynamical correction proportional to $v_{\perp}/c.$

 We now turn to a rigorous computation of the phase with the influence functional method~\cite{FeynmanVernon}, which allows us to capture non-local effects. The monitoring of only a subset of the d.o.f.s - the external atomic motion - calls for a partial trace (or coarse-graining) over the EM field and dipole moment. The time evolution of the reduced density matrix for the external atomic d.o.f. is obtained from 
  closed time path (CTP) integrals~\cite{CalzettaHu}. Such path integrals
 involve simultaneously forward and backward histories of the system.  
The coarse-graining over the environment d.o.f.s yields an influence functional connecting these two histories, which will be associated to a pair of interferometer paths. The influence of the environment onto the external atomic waves, captured by the influence functional, is at the origin of a path entanglement inducing the double-path phase discussed below.  
 
 First, we specify the actions for the considered quantum d.o.f., namely $S_{E}[\mathbf{r}_a] = \int_{0}^{T} dt   \: \left( \frac {m} {2} \dot{\mathbf{r}}_a^2(t) - V [\mathbf{r}_a(t)] \right)$ for the atomic position $\mathbf{r}_a$, $S_F[A^{\mu}] = (\epsilon_0 / 4)  \int  d^4 x \: F^{\mu \nu} F_{\mu \nu} $  with $F_{\mu \nu} \! = \! \partial_{\mu} A_{\nu}- \partial_{\nu} A_{\mu}$ for the EM field, and the action for the dipolar interaction
 $S_{AF}[A_{\mu},\mathbf{d},\mathbf{r}_a] = - \int  d^4 x \: J^{\mu}(x) \: A_{\mu}(x) \nonumber$ defined in terms of the current
 $J^{\mu}[\mathbf{d},\mathbf{r}_a](x) = - \int d t \:\sum_{i} d_{i}(t) \: \kappa^{\mu}_i \: \delta^{4} \left( x - 
 r_a(t) \frac {} {} \right)$. The four-dimensional integrals are defined as $\int d^4x \! \equiv \! \int_{0}^{T} dt \int d^3 \mathbf{r}$
 and the four-vector  $ r_a(t)\equiv({\bf r}_a(t),t).$ We have introduced the differential operator $\kappa_{i \mu} \! = \! \partial_i \eta_ {0  \mu}- \partial_0 \eta_{i  \mu} $ ($\eta_{\mu \nu}$ is the Minkowski metric with mostly plus signature) relating the electric field to the vector potential $A_{\mu}$
  by contraction: $E_i(x)= \kappa^{\mu}_i A_\mu(x)$. 
  The following discussion is valid for an arbitrary action $S_D[\mathbf{d}]$ for the internal atomic d.o.f.. We consider the evolution of the density matrix 
  $\rho(\mathbf{r}_a, \mathbf{r}_a', \mathbf{d}, \mathbf{d}',A^{\mu},A^{\mu}{}';t)$
   and trace over the field and dipole d.o.f.. The result can be expressed as a CTP integral over the position involving an influence action $S_{IF}[\mathbf{r}_a,\mathbf{r}_a']$
\begin{eqnarray}
\rho(\mathbf{r}_f,\mathbf{r}_f';T \!)  =  \int_{\rm CTP}^{\mathbf{r}_f,\mathbf{r}_f'} \!\! \mathcal D \mathbf{r}_a  e^{\frac {i} {\hbar} \left( S_E[\mathbf{r}_a] - S_E[\mathbf{r}_a'] + S_{IF}[\mathbf{r}_a,\mathbf{r}_a'] \right)}
\end{eqnarray}
where $\mathbf{r}_a$ and $\mathbf{r}'_a$ refer to forward and backward histories, respectively. 
We use the compact notation for the CTP 
 integral over a generic d.o.f. $X$
\begin{equation}
 \int_{\rm CTP}^{X_f,X'_f}  \! \! \!  \mathcal D X \!  = \!  \int \! d X_0 d X'_0 \int_{X_0}^{X_f} \! \mathcal D X \! \int_{X'_0}^{X'_f}  \! \mathcal D X' \! \rho \left( X_0,X'_0;0 \right)
\end{equation}
 with $\rho(0)$ the initial density matrix. 
 
 At this level, the influence action $S_{IF}[\mathbf{r}_a,\mathbf{r}_a']$ fully accounts for the effects of the field and the internal atomic dynamics on the external atomic d.o.f.. 
 As in the derivation of the local phase, we now assume that the wave-packets propagating along the two interferometer arms  are very narrow.
 The final density matrix $\rho(T)$ then contains four sharp peaks in the region $(\mathbf{r}_a,\mathbf{r}_a')$ centered around the classical positions $(\mathbf{r}_{k}(T),\mathbf{r}_{l}(T))$ for $k,l = 1,2$. The desired phase shift is obtained from the off-diagonal density matrix elements (i.e. $k \neq l$). 
 The main contribution to the path integral comes from the paths in the vicinity of the two stationary paths. Thus, the phase difference due to quantum fluctuations can be evaluated on the classical paths by taking the real part of the influence action $\phi_{if}=\frac {1} {\hbar}\mbox{Re}\left[ S_{IF}[\mathbf{r}_{1},\mathbf{r}_{2}] \right]$~\cite{Petruccione} (whereas the imaginary part represents the decoherence due to the plate~\cite{CasimirDecoherence}). 
 Note that the influence action $ S_{IF}[\mathbf{r}_{1},\mathbf{r}_{2}]$ depends in general on the entire paths followed from $t=0$ to $t=T,$
  and not only on the end-points.

Let us detail the procedure to obtain the influence action in the spirit of previous derivations of non-equilibrium forces mediated by 
a quantum field~\cite{Ryan2, Ryan1011}. First, we define an intermediate influence action 
$S_{IF}^{A_\mu }[\mathbf{d},\mathbf{d}',\mathbf{r}_a,\mathbf{r}_a']$ corresponding to the trace over the EM field alone: 
\begin{eqnarray}
\label{eq:influence functional field only}
& & e^{\frac i \hbar S_{IF}^{A_\mu }[\mathbf{d},\mathbf{d}'\!,\mathbf{r}_a,\mathbf{r}_a'\!]}  \! =   \! \int  \! \mathcal D A^f_{\mu}  \int_{\rm CTP}^{A^f_{\mu},A^f_{\mu}}  \! \mathcal D A_{\mu} \nonumber \\
& & \times e^{\frac i \hbar \left( S_F[A_\mu]+S_{AF}[A_{\mu},\mathbf{d},\mathbf{r}_a] -  S_F[A'_\mu]  - S_{AF}[A'_{\mu},\mathbf{d}',\mathbf{r}_a']  \right) }
\end{eqnarray}
This CTP integral over the vector potential linearly coupled to an external current yields
\begin{eqnarray}
\label{eq:influence functionalA}
S_{IF}^{A_{\mu}}[\mathbf{d},\mathbf{d}',\mathbf{r}_a,\mathbf{r}_a'] \!  & =&  \! \int \! d^4 x  d^4 x' \Bigl[ J^{\mu-} (x) G_{\hat{A},  \mu \nu}^{R}(x,x'\!) J^{\nu+} (x'\!)  \nonumber \\
&&  + \frac {i} {4} J^{\mu-} (x) G_{\hat{A}, \mu \nu}^H(x,x') J^{\nu-} (x') \Bigr] 
\end{eqnarray}
Taking standard conventions, we have introduced the semi-sum $J^{\mu +} \! = \! \frac 1 2  J^{\mu}[\mathbf{d},\mathbf{r}_a]+ \frac 1 2 J^{\mu}[\mathbf{d}',\mathbf{r}_a']$ and difference $J^{\mu -} \! = \! J^{\mu}[\mathbf{d},\mathbf{r}_a]-J^{\mu}[\mathbf{d}',\mathbf{r}_a']$ variables. The vector potential's retarded and Hadamard Green's functions  $G_{\hat{A}, \mu \nu}^{R,H}$ are defined as the electric field Green's functions 
$\mbox{G}_{\hat{\mathbf{E}}, ij}^{R,H}$ in Eqs.~(\ref{eq:def Green functions})-(\ref{Hadamard}) with Cartesian coordinates replaced by Lorentz indices. 

To obtain the desired influence action $S_{IF}[\mathbf{r}_a,\mathbf{r}_a']$, we average the EM influence functional given by
 Eqs.~(\ref{eq:influence functional field only},\ref{eq:influence functionalA}): $e^{\frac {i} {\hbar}  S_{IF}[\mathbf{r}_a,\mathbf{r}_a']}   =  \langle e^{\frac i \hbar S_{IF}^{A_\mu }[\mathbf{d},\mathbf{d}',\mathbf{r}_a,\mathbf{r}_a']  } \rangle_{\mathbf{d}}$ with $\langle ... \rangle_{\mathbf{d}}$
 denoting
 the time-dependent average over the free-evolving dipole d.o.f.. As in the derivation of the local phase
 (\ref{eq:phase usual atom interferometry2}), we take the approximation of small dipolar coupling and expand the influence functionals to first order in the 
 atomic polarizability: 
 $e^{\frac {i} {\hbar}  S^{A_{\mu}}_{IF}}\approx 1+\frac i \hbar  S^{A_{\mu}}_{IF}$ and likewise for 
   $e^{\frac {i} {\hbar} S_{IF}}.$ 
  The influence action then reads $S_{IF}[\mathbf{r}_a,\mathbf{r}_a'] \simeq \langle S_{IF}^{A_\mu }[\mathbf{d},\mathbf{d}'\!,\mathbf{r}_a,\mathbf{r}_a'] \rangle_{\mathbf{d}}$ \cite{Ryan2}.  In order to express the influence action in terms of
  EM and dipole correlation functions,  we  expand the currents $J^{\mu \pm}$ in 
   Eq.~(\ref{eq:influence functionalA}) and integrate over the spatial coordinates. The influence action receives a single-path (SP) and a double-path (DP) contribution, i.e. $S_{IF}   = S_{IF}^{\rm SP}+S_{IF}^{\rm DP} $ with 
\begin{eqnarray}
\label{eq:influence functional complex phase shift}
& &  S_{IF}^{\rm SP}[\mathbf{r}_{1},\mathbf{r}_{2}] = \! \frac {\hbar} {2} \mathop{\int \!\!\! \int_0^T}\! dt dt' \Bigl\{  \frac {} {} g^F_{t,t'}  \mathcal{G}_{\hat{\mathbf{E}}}^{R} \! \left( r_{1}(t),r_{1}(t')\right)  \nonumber \\
&  \:   & \qquad \qquad  \qquad \qquad \qquad - g^{F *}_{t,t'}  \mathcal{G}_{\hat{\mathbf{E}}}^{R} \! \left( r_2(t),r_{2}(t')\right)   \label{SPfinal} \\ 
  & + & \! \frac {i} {2}   \left[ \frac {} {} \! g^F_{t,t'}  \mathcal{G}_{\hat{\mathbf{E}}}^{H} \! \left(r_{1}(t),r_{1}(t')\right) 
    \! + \!   g^{F *}_{t,t'}   \mathcal{G}_{\hat{\mathbf{E}}}^{H} \! \left( r_{2}(t),r_{2}(t')\right)  \right] \Bigr\} \nonumber \\
   & &  S_{IF}^{\rm DP}[\mathbf{r}_{1},\mathbf{r}_{2}] =   \! \frac {\hbar} {2}  \mathop{\int \!\!\! \int_0^T}\! dt dt' \left\{ \frac {} {} g_{t',t}    \mathcal{G}_{\hat{\mathbf{E}}}^{R} \! \left(  r_{1}(t),r_{2}(t')  \right) \right. \nonumber \\
     &  \:   & \qquad \qquad\qquad\qquad\qquad
       -  g^{*}_{t',t}  \mathcal{G}_{\hat{\mathbf{E}}}^{R} \! \left(  r_{2}(t),r_{1}(t') \right) \frac {} {}  \label{DPfinal} \\
& - &  \frac {i} {2} \left. \left[ \! g_{t',t}   \mathcal{G}_{\hat{\mathbf{E}}}^{H} \! \left( r_{1}(t),r_{2}(t'\!)\right)   +  g^{*}_{t',t} \mathcal{G}_{\hat{\mathbf{E}}}^{H} \! \left( r_{2}(t),r_{1}(t')\right) \frac {} {} \! \right] \! \right\} \nonumber 
\end{eqnarray}
The plain and  time-ordered correlations of a free-evolving  dipole are defined as 
 $g_{t,t'}= \frac {1} {\hbar} \langle \hat{d}_i(t) \hat{d}_i(t') \rangle_{\mathbf{d}}$ and $g^F_{t,t'}= \frac {1} {\hbar} \langle {\cal T} [\hat{d}_i(t) \hat{d}_i(t')] \rangle_{\mathbf{d}},$ repectively  (Cartesian index $i$ omitted by isotropy).

To evaluate the SP influence phase
 $\phi^{\rm SP}_{if}=\frac {1} {\hbar} \mbox{Re}\left[ S_{IF}^{\rm SP}[\mathbf{r}_{1},\mathbf{r}_{2}] \right],$
  we note that both electric field Green's functions are real and use the general relations
    $\mbox{Re}\left[ g^F_{t,t'} \right]\!=\!\frac 1 2 G^H_{\hat{d}}(t,t')$ and $\mbox{Im}\left[ g^F_{t,t'} \right]=-\frac {1}{2} \left( G^R_{\hat{d}}(t,t')+G^R_{\hat{d}}(t',t)  \right)$. One then obtains from (\ref{SPfinal}) that $\phi^{\rm SP}_{if}=\phi_{loc}^{(1)}-\phi_{loc}^{(2)}$, namely
 the SP influence phase coincides exactly with the  local phase given by  Eq.~(\ref{eq:phase usual atom interferometry2}).
Thus, we recover the standard $ABCD$ van der Waals (Casimir-Polder for long distances) phase for long interaction times in the case of a constant atom-surface distance.

On the other hand, the presence of a non-local double-path phase  $\phi^{\rm DP}_{if}= \frac {1}
{\hbar} \mbox{Re}\left[ S_{IF}^{\rm DP}[\mathbf{r}_{1},\mathbf{r}_{2}] \right]$  contrasts sharply
with the local phase obtained by a standard atom interferometric method.
Eq.~(\ref{DPfinal}) shows that $\phi^{\rm DP}_{if}$  depends jointly on both interferometer paths and cannot be split into separate contributions associated to  individual arms. 

The contribution of the  
 Hadamard Green's function $ \mbox{G}_{\hat{\mathbf{E}}}^{H}$ in Eq.~(\ref{DPfinal})
 oscillates very rapidly as a function of $T$ and hence
 can be neglected in practice. Thus, the DP phase $\phi^{\rm DP}_{if}$ is the sum of the contributions from the 
 free-space [$\mathcal{G}_{\hat{\mathbf{E}}}^{R, 0}(x,x')$] and the scattering [$\mathcal{G}_{\hat{\mathbf{E}}}^{R, S}(x,x')$, Eq.(\ref{G_sca})] electric field retarded Green's functions, noted $\phi^{\rm DP}_{0}$ and $\phi^{\rm DP}_{S}$ respectively. Since the free-space DP phase $\phi^{\rm DP}_{0}$ is much smaller and independent of the distance $z_0$, we focus on the scattering DP phase $\phi^{\rm DP}_S$ capturing the plate influence:
 \begin{eqnarray} 
 \label{phiS}
 \phi^{\rm DP}_{S} = &  \! \frac {1} {2} \int_{0}^{T} dt\, \int_0^{t}dt' \, \mbox{Re}(g_{t',t}) \! &
\! \Bigl[  \!  {\cal G}_{\hat{\mathbf{E}}}^{R,S} \! \left(  r_{1}(t),r_{2}(t')  \right)  \\
  &&  -{\cal G}_{\hat{\mathbf{E}}}^{R,S} \! \left(  r_{2}(t),r_{1}(t')  \right) \Bigr] \nonumber 
 \end{eqnarray}

The phase given by Eq.~(\ref{phiS}) appears as the difference between the propagation integrals connecting the two separate arms,
which correspond to the 
 Feynman-like diagrams (b)-(c) in Fig.~1.
Each diagram represents the propagation from the retarded image four-position $r_{{\rm I} k}(t')$ to the 
 advanced 
four-position $r_l(t)$ corresponding to the other arm  ($l \neq k$) and with $t>t'.$ They express the electromagnetic interaction between the fluctuating dipoles of
 two coherent components corresponding to the quantum state of a single atom.

 In order to analyze these diagrams in more detail, we
  assume that the two arms in Fig.~1 share the same velocity component $v_{\parallel}$  parallel to the plate
during the time interval $T:$ 
   $\mathbf{r}_{1}(t)=v_{\|} t\,\hat{\mathbf{x}}+z_0\,\hat{\mathbf{z}}$ and 
 $\mathbf{r}_{2}(t)=v_{\|}t\, \hat{\mathbf{x}}+(z_0+v_{\bot}t)\, \hat{\mathbf{z}}.$ 
  Let us first discuss the diagram (b). Since the propagation time
$\tau=t-t'=|\mathbf{r}_1(t)-\mathbf{r}_{{\rm I} 2}(t')|/c$
is constrained by the speed of light [see Eq.~(\ref{G_sca})], 
the wave-packet  on the advanced arm 1 moves ahead of the image of the wave-packet on arm 2 during the  interval
 $\tau.$ Likewise, the wave-packet on arm 2 moves ahead of the image of arm 1 in diagram (c). 
 In the non-relativistic limit, $\tau$ is the same for both diagrams, and so is the displacement parallel to the plate during $\tau.$ 
 However, the  distance 
 $|\mathbf{r}_2(t)-\mathbf{r}_{{\rm I} 1}(t')|$ (c) is larger than $|\mathbf{r}_1(t)-\mathbf{r}_{{\rm I} 2}(t')|$ (b) because the wave-packet in arm 2 is moving away from the surface. 
 Since the retarded Green's function decreases as a function of distance, 
  diagram (b) corresponds to a stronger cross-talk between the quantum dipole fluctuations, leading to  
a positive  phase $\phi^{\rm DP}_{S}$ in Eq.~(\ref{phiS}). Thus, the DP phase is essentially a signature of the asymmetry
 between  diagrams (b) and (c), which is
 brought into play by the combination of two properties:
 \begin{itemize}
 \item
 The finite speed of the propagation. Note that 
 $\phi^{\rm DP}_{S}$ vanishes when  $c\rightarrow\infty.$ In fact,  $t'\rightarrow t$  leads 
 to an exact cancelation between the two diagrams since 
$|\mathbf{r}_2(t)-\mathbf{r}_{{\rm I} 1}(t)| = |\mathbf{r}_1(t)-\mathbf{r}_{{\rm I} 2}(t)|$  (this limit corresponds to vertical  
purple lines in Figs~1b and 1c). 
\item
The large memory time of dipole fluctuations. If the dipole correlation times were shorter than $z_0/c,$
each separate diagram contribution would be suppressed after multiplying  by 
$g_{t',t}$ in Eq.~(\ref{phiS}). In other words, the dipole memory time should be sufficiently large to enable the electromagnetic cross-talk between one arm and 
the image of the other arm. More generally, very short-living fluctuations 
lead to a coarse-grained evolution with no coupling between forward and backward histories of the system. Our double-path phase is precisely the 
signature of such a coupling for the atomic center-of-mass evolution.
 
 \end{itemize}

According to our convention, positive values for the double-path phase $\phi^{\rm DP}_{S}$ have the same  interferometric  effect  of 
a standard local phase on arm 1 larger than on arm 2. For the paths shown in Fig.~1, $\phi^{\rm DP}_{S}$ adds to the 
 van der Waals local phase difference 
 since arm 1 is closer to the plate than arm 2. However, the sign of $\phi^{\rm DP}_{S}$ is not determined by which path is closer (in average) to the plate, but rather 
 by which path 
  is moving away/towards the plate.  For instance, 
 $\phi^{\rm DP}_{S}$ would be negative if path 2 were moving towards the plate.

In order to derive simple analytical results, 
    we model the internal atomic d.o.f.
   as an harmonic oscillator with a transition frequency $\omega_0.$  The dipole correlation function is then proportional to the static atomic polarizability
    $\alpha(0):$
   $g_{t,t'}=\frac 1 2 \alpha(0) \omega_0 e^{- i \omega_0(t-t')}$ 
 (the frequency dependent polarizability $\alpha(\omega)$ is the Fourier transform of the dipole retarded Green's function $G^R_{\hat{d}}(\tau)\equiv 
 G^R_{\hat{d}}(t'+\tau,t')$). 
  In the short-distance limit $z_0,  v_{\perp} T \ll \lambda_0$, Eq.~(\ref{phiS}) leads to
\begin{equation} \label{final}
\phi^{\rm DP }_{S}= \frac {3 \pi} {4 \lambda_0} \left( \frac {\alpha(0)} {4 \pi \epsilon_0} \right)  \left( \frac {1}  {z_0^2} - \frac {1}  {(z_0+ v_{\perp} T/2)^2} \right)
\end{equation}
where $\lambda_0=2 \pi c / \omega_0$ is the transition wavelength.
For a long path separation, i.e. $z_0 \ll v_{\perp} T \ll  \lambda_0$, the DP phase saturates to a maximal value independent of $v_{\perp} T.$
We compute the saturation value for $\:^{87} \mbox{Rb}$ atoms, with the 
static polarizability $\alpha_{\rm Rb}(0)/(4 \pi \epsilon_0) =4.72 \times 10^{-29} \: \mbox{m}^3.$ 
The dominant contribution to the ground state dipole fluctuations 
 comes from the $5s_{1/2}-5p_{1/2}$ and  $5s_{1/2}-5p_{3/2}$ transitions, with
wavelengths close to
$\lambda_0 \simeq 0.79 \: \mu \mbox{m}$ (they correspond to large dipole transition matrix elements).  
We take
 $z_0 = 20 \: \mbox{nm}$ similar to the distance used in the experiments of Ref.~\cite{CroninVigue}. Such parameters yield a DP phase of $\phi^{\rm DP}_{if}= 3.5 \times 10^{-7} \: \mbox{rad}$, hence beyond the sensitivity given by the state of the art  atom interferometers but still larger than systematic phases considered in atom gravimeters~\cite{Kasevich07}. This DP phase can be compared with the standard vdW phase $\phi^{(2)}_{\rm vdW}$ obtained by integration of the potential $V_{\rm vdW}(z)=-\hbar \omega_0 \alpha(0)/(32 \pi\epsilon_0 z^3)$ along the path $2$. The latter is inversely proportional to the normal velocity $v_\perp$ and reads $\phi^{(2)}_{\rm vdW}= ( \frac {c} {v_\perp}) \left( \frac { \pi} {8 \lambda_0} \right) \left( \frac {\alpha(0)} {4 \pi \epsilon_0} \right) \left( \frac {1}  {z_0^2} - \frac {1}  {(z_0+ v_{\perp} T)^2} \right) $. In the considered limit $\phi^{\rm DP}_{if} \simeq 6 (v_\perp/c) \phi^{(2)}_{\rm vdW}$, showing that the non-local DP phase is a first-order relativistic correction to the standard vdW phase, in agreement with our discussion about the role of the finite value of $c.$ This explains why the DP phase discussed here is several orders of magnitude smaller than the typical vdW phase $\phi_{\rm vdW} \sim 0.2 \: \mbox{rad}$ measured in Refs.\cite{CroninVigue}. It is actually hard to isolate the non-local DP phase from the much larger standard vdW phase in the atom interferometer discussed here. Other interferometer setups better suited for that purpose remain to be investigated.

To conclude, we have developed an open quantum system theory of atom interferometers, predicting non-local double-path phase shifts in the propagation of atomic waves. We have shown that the standard atom-optics approach catches only the local phase shifts, which correspond to the single-path terms obtained with the influence functional method. 
The atomic center-of-mass is coupled to  dipole and EM fluctuations which play the role of a common environment for the two wavepackets propagating in the interferometer. The coherence of matter waves and large dipole memory times allow for cross-talks between the dipole fluctuations on each arm, 
leading to the non-local DP phase. 
We have developed a diagrammatic picture of this
one-particle quantum interference effect, which can be interpreted as an asymmetry between diagrams involving  simultaneous atomic propagation on distinct paths.
The finite speed of light allows quantum fluctuations to probe this asymmetry. Thus, the DP phase can be interpreted as a dynamical relativistic correction.
We have shown that the DP phase shift compares to systematics considered for accurate atom interferometers. Our approach can be extended to multiple-path atom interferometers by considering pairs of paths.This is to our knowledge the first evidence of a non-local phase coherence in atom optics.

\acknowledgments

The authors are grateful to Christian J. Bord\'e, Diego A. R. Dalvit, Arnaud Landragin, and Reinaldo de Melo e Souza for stimulating discussions. This work was partially funded  by CNRS (France), CNPq , FAPERJ and CAPES (Brasil).

\end{document}